\documentclass{article}
\usepackage{arxiv}

\RequirePackage{fix-cm}
\usepackage{graphicx}
\newcommand{\shellcmd}[1]{\indent\indent\texttt{\footnotesize\# #1}}

\usepackage{natbib}

\usepackage{booktabs}
\usepackage{makecell}
\usepackage{adjustbox}
\newcolumntype{R}[2]{%
    >{\adjustbox{angle=#1,lap=\width-(#2)}\bgroup}%
    l%
    <{\egroup}%
}
\newcommand*\rot{\multicolumn{1}{R{90}{-0.1em}}}
\newcommand*\rottwo{\multicolumn{1}{R{90}{2.35em}}}

\usepackage{colortbl}
\usepackage{pdflscape}
\usepackage{afterpage}
\usepackage{capt-of}

\usepackage{xcolor}

\definecolor{mycolor}{rgb}{0,0,128}
\newcommand\hl[1]{%
  \bgroup
  \hskip0pt\color{black!80!black}%
  #1%
  \egroup
}

\usepackage[colorlinks,citecolor=black,urlcolor=black]{hyperref}

\usepackage{tikz}
\def\checkmark{\tikz\fill[scale=0.4](0,.35) -- (.25,0) -- (1,.7) -- (.25,.15) -- cycle;} 
\usepackage[switch]{lineno}

\begin{document}

\title{Experiments as Code: \\
\large A Concept for Reproducible, Auditable, Debuggable, Reusable, \& Scalable Experiments
}

\author{
Leonel Aguilar$^{1,2,\dagger}$ \href{https://orcid.org/0000-0001-6864-4492}{\includegraphics[scale=0.09]{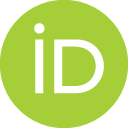}}\and
Michal Gath-Morad$^{1,\dagger}$ \href{https://orcid.org/0000-0001-7673-6290}{\includegraphics[scale=0.09]{figures/orcid.png}}\and	
Jascha Gr\"ubel$^{1,3,\dagger}$  \href{https://orcid.org/0000-0002-6428-4685}{\includegraphics[scale=0.09]{figures/orcid.png}}\and
Jasper Ermatinger$^1$ \and
Hantao Zhao$^{1}$ \href{https://orcid.org/0000-0003-0398-3842}{\includegraphics[scale=0.09]{figures/orcid.png}}\and
Stefan Wehrli$^5$ \href{https://orcid.org/0000-0001-9845-9028}{\includegraphics[scale=0.09]{figures/orcid.png}}\and
Robert W. Sumner$^3$
\href{https://orcid.org/0000-0002-1909-8082}{\includegraphics[scale=0.09]{figures/orcid.png}} \and
Ce Zhang$^{2}$\and
Dirk Helbing$^5,^6$ \and
Christoph H\"olscher$^{1,5}$
\href{https://orcid.org/0000-0002-5536-6582}{\includegraphics[scale=0.09]{figures/orcid.png}}
}

\maketitle

\begin{abstract}
A common concern in experimental research is the auditability and reproducibility of experiments.
Experiments are usually designed, provisioned, managed, and analyzed by diverse teams of specialists (e.g., researchers, technicians and engineers) and may require many resources (e.g. cloud infrastructure, specialized equipment).
Even though researchers strive to document experiments accurately, this process is often lacking, making it hard to reproduce them.
Moreover, when it is necessary to create a similar experiment, very often we end up “reinventing the wheel” as it is easier to start from scratch than trying to reuse existing work, thus losing valuable embedded best practices and previous experiences.
In behavioral studies this has contributed to the reproducibility crisis.
To tackle this challenge, we propose the ``Experiments as Code'' paradigm, where the whole experiment is not only documented but additionally the automation code to provision, deploy, manage, and analyze it is provided.
To this end we define the Experiments as Code concept, provide a taxonomy for the components of a practical implementation, and provide a proof of concept with a simple desktop VR experiment that showcases the benefits of its ``as code'' representation, i.e., reproducibility, auditability, debuggability, reusability, \& scalability.

\end{abstract}
\vspace{3cm}
Contact: Leonel Aguilar\\ leonel.aguilar@gess.ethz.ch\\
   $^1$ Chair of Cognitive Science, ETH Z\"urich\\
   $^2$ Data Science, Systems and Services Group, ETH Z\"urich\\
   $^3$ Game Technology Center, ETH Z\"urich\\
   $^5$ Decision Science Laboratory, ETH Z\"urich\\
   $^6$ Chair of Computational Social Science, ETH Z\"rich\\
   $\dagger$ These authors provided equivalent contributions

\twocolumn

\section{Introduction}
\label{intro}

Over the last decade, we have observed a crisis of scientific reproducibility in behavioral experimental research \citep{ioannidis2005most,pashler2012editors,open2015estimating,camerer2018evaluating}. 
This reproducibility crisis has exposed several short-comings of how experiments are communicated within the scientific community and beyond.
The crisis is very much rooted in the fact that implicit information about the experimental process often gets communicated as \emph{viva voce} \citep{hassani2018communication}.
Third parties attempting to reproduce an experiment may fail due to missing such implicit information.

The COVID-19 pandemic is exacerbating the crisis of reproducibility due to the shift towards remote work and limiting in person experiments.
This, in turn, further limited the ability to communicate such information directly within a research context.
The diminished opportunities for interaction during the pandemic caused major side-and-after effects resulting in a decrease of experimental data quality and unnecessary extra costs within a research group and the research community at large.

As a way to achieve reproducibility, \cite{stark2018before} introduced the term ``Preproducibility''. 
Preproducibility, or scientific recipes, requires researchers to provide the complete set of necessary elements and procedures to replicate an experiment that ``cannot with advantage be omitted'' \citep[p.44]{popper1992open}.
In contrast, current efforts to achieve reproducibility tend to provide only sufficient information to \emph{understand} the experiment in detail. 
Understanding an experiment is not enough in the process of re-implementing the experiment from scratch.
Nuanced details may be implemented differently, which ultimately changes the outcome  \citep[e.g.,][]{reardon2016mouse}.
Any future research needs to address these concerns to underpin the validity of their claims or explain deviations from previous results like behavioral drifts or changes in implicit experimental conditions \citep[e.g.,][]{reardon2016mouse}.

The relatively new field of \emph{meta-science} \citep{ioannidis2015meta} and initiatives across all sciences \citep{paperWithCode,renku,aguilar2021ease,orchard2003proteomics, welter2014nhgri} aim to improve experimental research \citep{baumgaertner2018openness} and enable both reusability and composability \citep{nierstrasz1995research}.
However, these important initiatives and suggestions disjunctively solve partial problems. 
To bridge this gap, we require a holistic approach that ties together the benefits of these initiatives. 
Behavioral experiments in VR are a useful testing ground to adopt a more holistic approach because they pose unique reproducibility challenges and an opportunity to implement Preproducibility.
On the one hand, the susceptibility of human behavior to nuanced changes can drastically alter experiment outcomes \citep{kuliga2015virtual, Gath-Morad2020how}.
On the other, VR experiments are often fully virtual and offer themselves to automation and thus formal documentation \citep[e.g.,][]{grubel2016eve}.

As a holistic approach to the reproducibility of VR experiments and as a practical implementation of Preproducbility we propose the paradigm \emph{Experiments as Code} (ExaC) inspired by the cloud computing world and in particular the ``\underline{\hspace{0.5cm}} as Code'' and ``\underline{\hspace{0.5cm}} as a Service'' revolutions.
Given the ubiquity and accessibility of cloud computing services, researchers who use ExaC could deploy their experiments without cloud computing expertise.
This would allow the reuse of published working configurations, effectively transferring and inheriting implicit knowledge essential for reproducibility.
ExaC implements preproducibility in VR experiments by documenting all its elements, procedures, information, and implemented best practices through code.
This includes the whole experimental pipeline, protocol, analysis tools, VR software, and the configuration of the required computing infrastructure.
The current publishing standards and disjoint initiatives already encourage better practices in the experimental process.
Nevertheless, the ExaC paradigm ties them together and overcomes the limitations imposed by \emph{viva voce} knowledge transfer.
ExaC transforms the implicit elements and procedures required for reproducibility into explicitly documented and automated code procedures.
Additionally, versioning of the code allows keeping track not only of the final working configurations but also intermediate steps, failed attempts, and configuration testing.
These steps are all recorded and annotated with information concerning the decisions or criteria that lead to changes in the code base through a version control system.


The challenges of reproducibility are often specific to the experiment type, nature and requirements. For example for local experiments, APIs interfacing infrastructure with specialized hardware needs to be built, e.g., interfacing with skin conductance sensors, specialized immersive equipment or CAVE \citep{cruz1992cave} environments. For remote data collection, \cite{ma2018web} identified challenges in crowd-sourced remote immersive VR experiments, e.g., the heterogeneity of hardware, participants trying to trick the experiment, and participant pool size and composition. 
The Human Computer Interaction (HCI) community has documented several reproducibility challenges, e.g., \cite{ratcliffe2021remote} collected experiences and challenges of remote experimentation in HCI  research, while \cite{echtler2018open} documented and discusses the reproducibility crisis and the extent of use of open source and open science concepts in HCI. \cite{feger2019role} glances over the roles and challenges of HCI on research reproducibility, e.g., the incentives for making research reproducible and the potential role of gamification mechanisms. 
Concrete recommendations such as the ones provided for remote VR data collection by \citep{mottelson2021} could be embedded and actionable in an ExaC representation.
We argue that, although ExaC by itself does not make an experiment reproducible or solve the challenges to achieve reproducibility it provides a path to it through a way to document, automate, instantiate and transfer proven working solutions.


To showcase the features of an ExaC representation we provide a proof-of-concept implementation of a desktop VR wayfinding experiment.
Our Simple Experiment as Code Template (SimpleExaCT) repository enables the reproducibility of this experiment.
The implementation uses several ``\underline{\hspace{0.5cm}} as Code'' components along with an ad-hoc VR framework, SimpleVR.
However, SimpleExaCT's modular approach makes it compatible with other VR frameworks.
Consequently, we review existing frameworks in the light of the ExaC paradigm.
We envision a new third generation of VR frameworks that goes beyond VR hardware integration (first generation) and experiment design (second generation) towards conducting reproducible, auditable, debuggable, reusable and scalable experiments.

The main contributions of this paper are three-fold.
First, we contribute towards the ExaC paradigm by providing a definition and demonstrating it with the full experimental pipeline of a desktop VR behavioral experiment in the domain of Cognitive Science.
Second, we provide an open source proof of concept implementation of this Experiment as Code, to scaffold data collection in our online desktop VR experiment. This enables third parties to audit and reuse our experimental design and showcases the benefits of the ExaC representation.
Third, we provide a detailed comparison of notable VR frameworks and review them through the lens of ExaC emphasising their capacity to support Preproducibility.
Although we showcase the ExaC paradigm with a desktop VR experiment, we envision its potential for improving experimental research in general.

\section{Background, Definitions, and Related Work}

\subsection{Meta-science, Reproducibility and the ``\underline{\hspace{0.5cm}} as Code'' revolution}

The ExaC paradigm is rooted in three larger developments of the last decade.
First is the rise of the field of \emph{meta-science} \citep{ioannidis2015meta}, second are initiatives to overcome the reproducibility crisis, and third are developments in cloud computing such as the ``\underline{\hspace{0.5cm}} as Code'' and ``\underline{\hspace{0.5cm}} as a Service'' revolutions.

Meta-science tries to improve the processes of scientific work by making the scientific inquiry itself the object of their study. Suggestions include pre-registration \citep{van2016pre}, a complex systems approach \citep{wallot2018interaction}, replication as teaching \citep{grahe2012harnessing}, change in attitudes to p-values \citep{colquhoun2014investigation}, larger samples \citep{maxwell2015psychology}, raw data access, triangulation instead of replication \citep{munafo2018robust}, and standards of presentation \citep{gosselin2020statistical}.

There are several initiatives that tackle challenges impacting directly reproducibility.
For instance, there is an indexing service that specializes in indexing papers accompanied by code \citep{paperWithCode}.
Another initiative provides multidisciplinary, collaborative and reproducible data pipelines \citep{renku}.
There are also modular ecosystems for improving machine learning reusability by managing and automating the complete machine learning application development life-cycle with reusable models and components \citep{aguilar2021ease}.
Some disciplines have pushed even further, creating domain specific standards \citep{orchard2003proteomics} and ontological databases \citep{welter2014nhgri} that enable more efficient and systematic knowledge extraction.
Moreover, publishers are now encouraging the publication of protocols in specialized services such as protocols.io, where protocols are converted to modifiable and reusable checklists.

Tightly interrelated with reproducibility are the notions of auditability, debuggability, reusability, \& scalability.
We define auditability as providing a transparent and comprehensive view into the experiments' mechanics, tools and components.
Thereby, any third party is able to evaluate the software \citep{weiss1980auditability} and by extension the experiment's reproducibility. We define debuggability as providing the means to identify and modify aspects of the experiment affecting its reproducibility \citep{rabinovich2014reproducibility}.
We define reusability as the ability to reuse parts or variations of an existing software to create a new one \citep{prieto1993status} and in our context software extends to experiments.
Here we can exploit the benefits of open source software development in science, \citep{echtler2018open}.
These three are ubiquitous tasks in open source software development and as such are enabled through the ExaC representation by inheriting the same good practices.
Finally, in systems scalability refers the ability to either expand a system in one dimension (structural) or an increase in the amount of activity (load) the system receives \citep{bondi2000characteristics}. In the context of experiments, this refers to the ability to increase the sample size either through repetition (horizontal) or the use of larger participant pools (vertical), e.g., crowd-sourced and remote experiments.

In cloud computing, the complexities and requirements of current web services have pushed advances and paradigm shifts such as the ``\underline{\hspace{0.5cm}} as Code'' and ``\underline{\hspace{0.5cm}} as a Service'' revolutions.
Paradigms such as ``Infrastructure as Code'' (IaC; e.g., Terraform, AWS Cloud formation), ``Environments as Code'' (e.g., Vagrant, Docker), ``Configuration as Code'' (e.g., Ansible, Puppet Chef), ``Data Pipelines as Code'' (e.g., Airflow) are now ubiquitous in web service development.
There, a single service requires the deployment of hundreds to thousands of micro-services, their infrastructure to be dynamically provisioned, and their data to be continuously analysed.

\begin{figure*}[htpb!]
  \includegraphics[width=\textwidth]{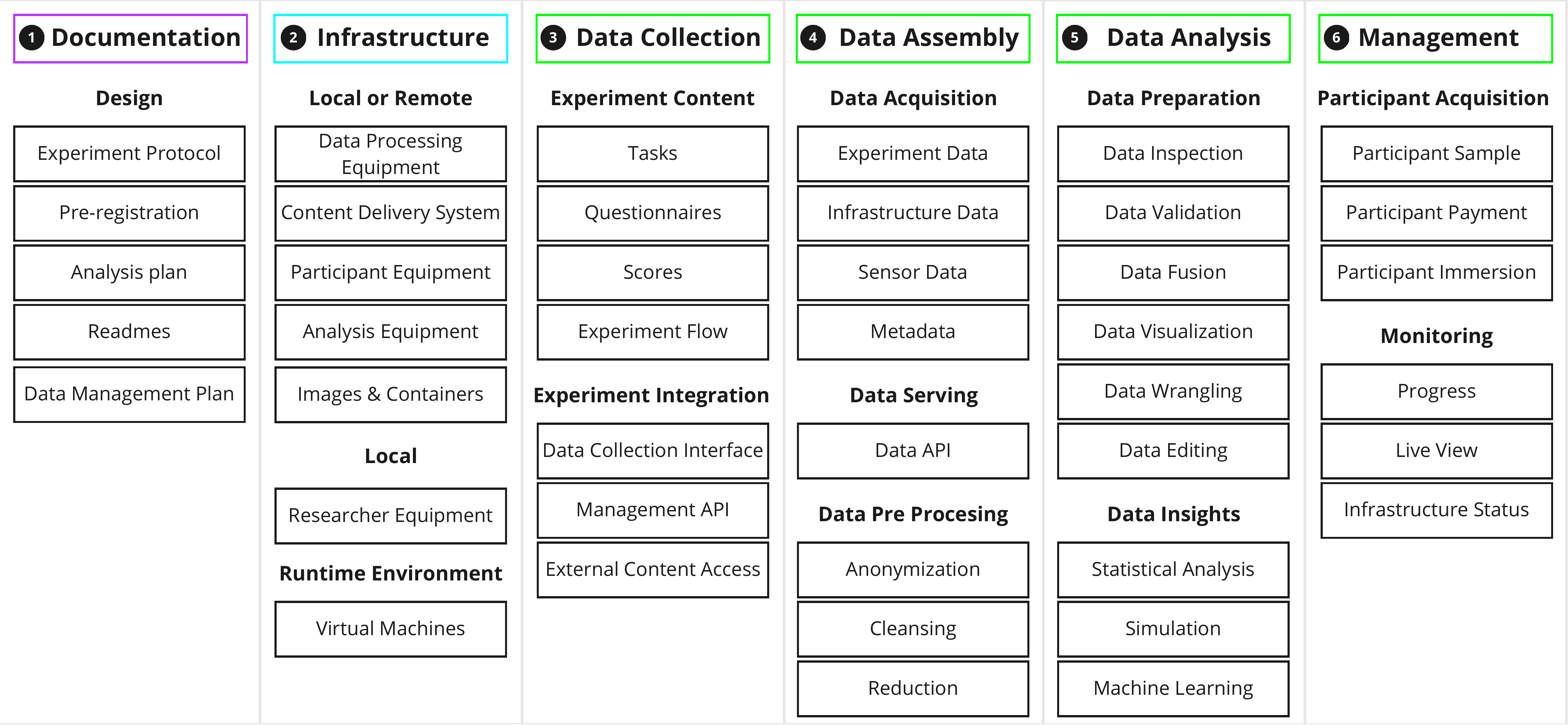}
\caption{Six pillars taxonomy of behavioral experimentation: 1) documentation, 2) infrastructure \& environment, 3) data collection, 4) data assembly, 5) data analysis, 6) management. Each pillar consists of multiple essential tasks that need to be completed. Each pillar is represented distinctively in the ExaC paradigm as a block for automation of an experiment.}
\label{fig:six:pillars}       
\end{figure*}

For example, IaC allows storing and versioning the infrastructure and configuration requirements of complex web-services.
These code representations have enabled simpler deployment and provisioning, rolling back to previous configurations, mirroring services, planning, scaling, and even debugging problems in the infrastructure provisioning.
We believe that many of the features of the ``\underline{\hspace{0.5cm}} as Code'' movement and their tools could enhance experiments in general and VR experiments in particular.

Despite this substantial drive for improving the reproducibility of behavioral experiments, Preproducibility is still a challenge. 
However, the boom of ``\underline{\hspace{0.5cm}} as Code'' provides a foundation to implement Preproducibility and achieve reproducibility in behavioral experiments.
To the best of our knowledge, the term ``Experiment(s) as Code'' has only been used in two research articles.
First, in EnosStack \citep{cherrueau2018enosstack} that proposes a software stack composed of Python, Ansible, Docker and their library EnosLib to enable reproducible experimental workflows.
Second and similarly, \cite{almaatouq2020empirica} identified the need for reproducible virtual labs and created a full platform.
In contrast to these works, we define the ExaC paradigm and scaffold it on to a taxonomy describing how to conduct VR experiments as an implementation of Preproducibility.
This entails combining a wide variety of approaches from ``\underline{\hspace{0.5cm}} as Code'' with VR frameworks and other complementary software components of the experiment.
Moreover, the components are only loosely coupled allowing to exchange them as services evolve.

\subsection{Behavioral experiments in VR and ExaC definition}

Behavioral experiments in VR can be composed of several elements (e.g., tasks, procedures, information, and documentation) beyond the experiment content.
They can span from its pre-registration over the acquisition of participants to the monitoring of the infrastructure status.
We classify these elements as components in groups and subgroups under the proposed six pillars taxonomy (see Fig.~\ref{fig:six:pillars}).
The pillars cover the essential aspects of an experiment by subdividing it into groups: documentation, infrastructure \& environment, data collection, data assembly, data analysis, and management.
We collect and exemplify these common components required to enable reproducibility.
The exemplified components may vary in their classification and extent depending on the details of their implementation and experiment requirements, e.g., sensitive data anonymization could be desirable to occur at collection time.
While some components of each pillar are strictly required to produce an experiment, others are optional but desired to improve reproducibility.

We base our definition of ExaC on the idea of preproducibility \citep{stark2018before}, but instead of providing \emph{scientific recipes} we aim at providing \emph{automated and digitally documented scientific recipes} inspired by the advances in cloud computing.
Thus, we define the ExaC paradigm as providing imperative, or declarative automation code and digital documentation to the components of the respective essential pillars. 

\begin{figure}[htpb!]
  \includegraphics[width=\columnwidth]{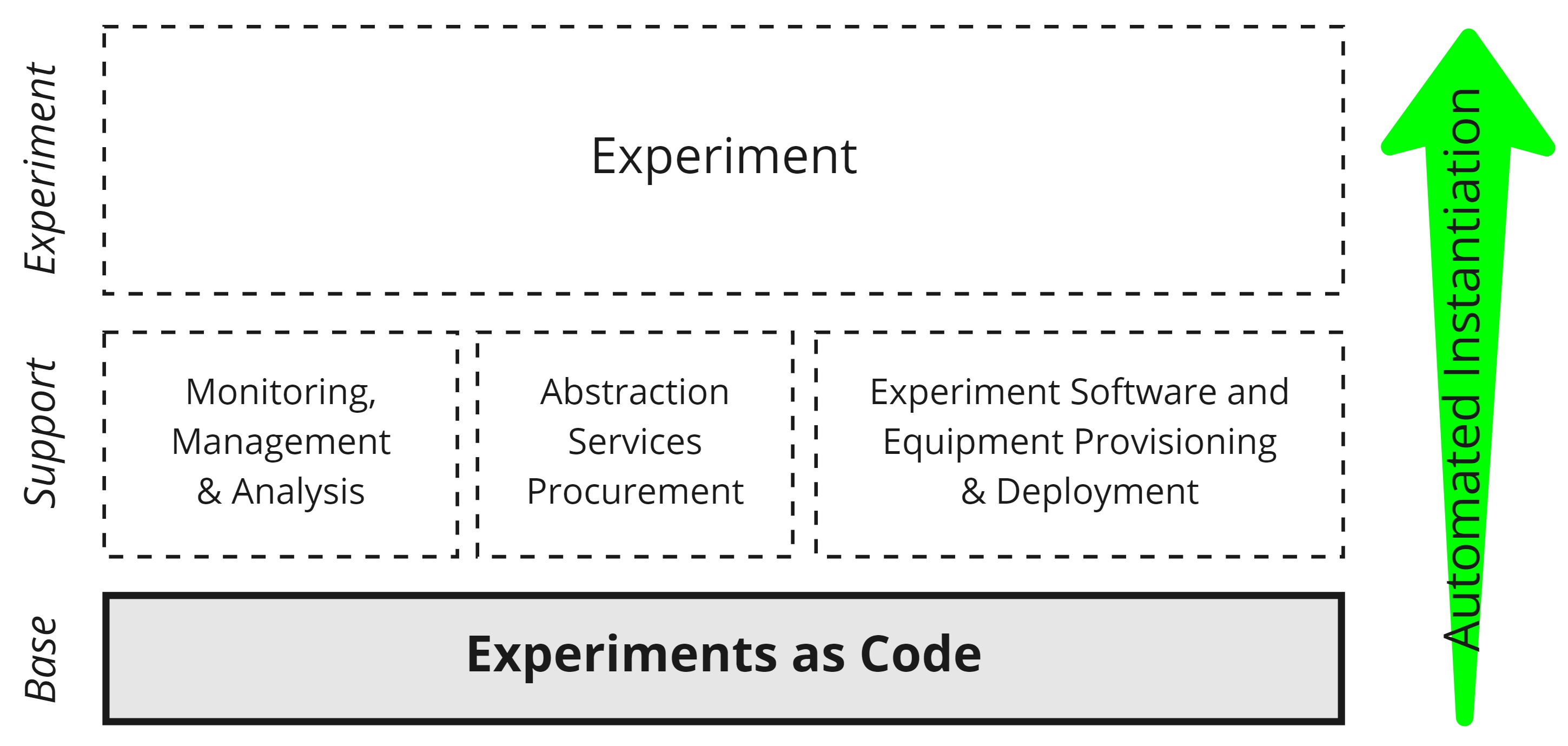}
\caption{ExaC paradigm application scheme: The paradigm serves as the Base image to instantiate the Support (infrastructure and services) and then conduct the Experiment itself. The ExaC (gray box) covers the 6 pillars taxonomy of behavioral experimentation (see Fig.~\ref{fig:six:pillars}).}
\label{fig:experiment:as:code:abstract}       
\end{figure}

We borrow the imaging concept from Environments as Code.
An image of how a run-time environment should look like serves as a specification of how actual containers are deployed.
The static ExaC code base serves as the image (see Fig \ref{fig:experiment:as:code:abstract}, Base).
The whole live experimental setup can be automatically instantiated from the image including the supporting components and the components building the experiment itself.
This entails but is not limited to provisioning the right hardware, deploying the software stack, instantiating abstraction interface services, acquiring participants, hardware monitoring services, serving the content to participants, and ultimately collecting the experiments data.

\subsection{VR Frameworks in the context of ExaC}

The pillars used in ExaC provide defined boundaries for modularization.
VR frameworks and specialized software can be used to provide automation code and digital documentation for them. 
Many existing VR Frameworks already provide support or potential for the automation of many of the common components described within the pillars.
We provide a summary of the features of these VR Frameworks through the lens of ExaC and the six pillars in Table \ref{tab:vr:frameworks:comparison}.
Additionally, we collect the documented efforts to address each of the specific components of the pillars in the Supplementary Information Section 1 and an in-depth discussion of the VR frameworks through the lens of ExaC in the Supplementary Information Section 2. 

\begin{table*}[htpb!]
    \centering
    \resizebox{\textwidth}{!}{%
    \begin{tabular}{lcc|cccc|cccc|cc|cc|cc|cc}
         \toprule
         &&\multicolumn{15}{c}{Experiments as Code}\\
         \cmidrule(rl){3-17}
         && \multicolumn{1}{c}{Doc.} &
         \multicolumn{4}{c}{\makecell{Infrastructure}} & \multicolumn{4}{c}{\makecell{Data  Collection}} & \multicolumn{2}{c}{\makecell{Data\\Assembly}} & \multicolumn{2}{c}{\makecell{Data\\Analysis}}& \multicolumn{2}{c}{Managem.} &\multicolumn{2}{c}{Access}\\
         \cmidrule(rl){3-3} \cmidrule(rl){4-7} \cmidrule(rl){8-11} \cmidrule(rl){12-13}  \cmidrule(rl){14-15} \cmidrule(rl){16-17} \cmidrule(rl){18-19}
         
         Framework & \rot{Platform} & \rot{{\color{gray}Protocol}$^\dagger$} & \rottwo{\makecell{Multi-user\\ Server}}& \rot{Immersive VR} & \rot{{\color{gray}Provisioning}$^\dagger$} & \rot{{\color{gray}Deployment}$^\dagger$} & \rot{Prefab. tasks} & \rottwo{\makecell{Task\\config. files}}& \rottwo{\makecell{Game engine\\code}} & \rot{Questionnaires} & \rot{Remote Storage}& \rot{Database} & \rot{Analysis} & \rot{Replay} & \rot{Participants} & \rot{Monitoring} &
         \rot{github/bitbucket} & \rot{Unity package}\\
         \midrule
         \makecell[l]{AGENT \\{\tiny\citep{moulec2017agent}}} & Any 
         &
         & & & \phantom{aai} & \phantom{aai}
         & & \checkmark & \checkmark & 
         & &
         & &
         & &
         \\
         
         \makecell[l]{bmlTUX \\{\tiny\citep{bebko2020bmltux}}} & Unity 
         &
         & & & &
         & & \checkmark & \checkmark & 
         & &
         & &
         & & \checkmark
         & \checkmark & 
         \\
         
         \makecell[l]{EVE \\{\tiny\citep{grubel2016eve}}} 
         & Unity 
         &
         & & * & &
         & \checkmark & * & \checkmark & \checkmark 
         &  & \checkmark
         & \checkmark & \checkmark
         & & 
         & \checkmark & 
         \\
         
         \makecell[l]{DeFINE\\{\tiny\citep{tiwari2021define}}}
         & Unity
         &
         & & \checkmark & &
         & & \checkmark & \checkmark & \checkmark
         & \checkmark & \checkmark
         & &
         & &
         & \checkmark &
         
         \\
         
         \makecell[l]{Landmarks \\{\tiny\citep{starrett2020landmarks}}} & Unity 
         &
         & & \checkmark & &
         & \checkmark & \checkmark & \checkmark &  \checkmark  
         & &
         & * &
         & &
         & \checkmark & \checkmark 
         \\
         
         \makecell[l]{NVR-DeSciL \\{\tiny\citep{zhao2018networked}}} & Unity 
         &
         & \checkmark & & &
         & & & \checkmark & 
         & \checkmark &
         & &
         & \checkmark & \checkmark 
         & &
         \\
         
         \makecell[l]{OpenMaze \\{\tiny\citep{alsbury2020openmaze}}} & Unity 
         &
         & & * & &
         &  & \checkmark & & 
         & &
         & &
         & &
         & \checkmark & 
         \\
         
         \makecell[l]{Toggle Toolkit \\{\tiny\citep{ugwitz2021toggle}}} & Unity 
         &
         &  & * &  &
         & * &  &  & 
         &   &  
         &   &
         &  & 
         & \checkmark &  
         \\
        
         \makecell[l]{VO \\{\tiny\citep{howie2020virtual}}} &  Unity 
         &
         & & \checkmark & &
         & & & \checkmark & 
         & \checkmark & \checkmark
         &  & \checkmark
         &  & 
         & &
         \\
         
         \makecell[l]{VRate \\{\tiny\citep{regal2018vrate}}}
         & Unity 
         &
         & & \checkmark & & 
         & & \checkmark & & \checkmark
         & &
         & &
         & &
         & & 
         \\
         
         \makecell[l]{VREVAL \\{\tiny\citep{schneider2018vreval}}} 
         & Unreal 
         & 
         & & \checkmark & &
         & \checkmark & & & \checkmark 
         & &
         & \checkmark &
         & &
         & & - 
         \\
         
         \makecell[l]{VREX\\{\tiny\citep{vasser2017vrex}}} 
         & Unity
         &
         & & \checkmark & &
         & * & * & * &
         & &
         & &
         & &
         & &
         \\
         
         \makecell[l]{VR-Rides \\{\tiny\citep{wang2020vr}}} & Unity  
         &
         & & \checkmark & &
         & & & \checkmark & 
         &   & \checkmark
         & * & 
         & &
         & \checkmark & 
         \\
         
         \makecell[l]{USE \\{\tiny\citep{watson2019use}}} & Unity 
         &
         & & & & 
         & & \checkmark & & 
         & &
         & &
         & &
         & \checkmark & 
         \\
         
         \makecell[l]{UXF \\{\tiny\citep{brookes2019studying}}} & Unity
         &
         & & \checkmark &  &
         & & \checkmark & \checkmark  & 
         & \checkmark & \checkmark
         & &
         & &
         & \checkmark & \checkmark
         \\
         \bottomrule
         Symbols: & \multicolumn{17}{r}{\checkmark: Feature, *: Partial Feature,  -: Not possible, $^\dagger$: Never implemented}
    \end{tabular}
    }
    \caption{Comparison of current VR frameworks since 2015. The features are grouped by correspondence to the 6 pillars taxonomy of the \emph{ExaC} paradigm. Features are evaluated on the description of the frameworks in their accompanying paper. We also report the availability of the frameworks.}
    \label{tab:vr:frameworks:comparison}
\end{table*}

In the early 2000s, a first set of VR frameworks \citep{bierbaum2001vr, tramberend1999avocado,ayaz2008maze, annett2009vr, mossel2012artifice} focused on simplifying the technical setup.
However, these frameworks were rapidly eclipsed by game engines (i.e., Unity and Unreal) and became quickly obsolete.
Over the last five years, a second generation of VR frameworks \citep{grubel2016eve,moulec2017agent,schneider2018vreval,zhao2018networked,brookes2019studying,watson2019use,alsbury2020openmaze,bebko2020bmltux,starrett2020landmarks,wang2020vr,ugwitz2021toggle} has put the experiment design at the core of the framework.
Technical issues such as rendering, physics and specialized hardware have been delegated to the underlying game engines such as Unity or Unreal and third-party libraries to integrate Human Interface Devices (HIDs; e.g. Oculus Touch, 3D mice).

The second generation of VR frameworks provide better support to the different components required for experimentation.
However, there is still a considerable gap in terms of hardware provisioning and deployment, support for data analysis, multi-user setups, and participant management and monitoring.
To bridge these gaps, we envision a third generation of VR frameworks that either natively support all six pillars of ExaC or that are modularly designed to better integrate with specialized tools that cover them. 
This third generation would extend their focus from experimental design towards reproducibility, auditability, debuggability, reusability, and scalability. 
General frameworks for behavioral experimentation recently started focusing on reproducibility \citep{cherrueau2018enosstack,almaatouq2020empirica}. 
However, we believe that for the future development of VR frameworks with reproducibility as the corner stone, a robust definition of the underlying paradigm to achieve and document reproducibility is required.
Hence this paper proposes and defines ExaC as the underlying paradigm and proposes a taxonomy to support its modular development.

\section{Method}
\label{sec:method}

To demonstrate the ExaC paradigm we provide as a proof-of-concept automation code covering each of the 6 pillars of experimentation (see Fig.~\ref{fig:six:pillars}).
For this, we used open source tools (i.e., terraform, jupyter-lab and docker) coupled with a minimal ad-hoc framework we named SimpleVR written in C\#/Unity, Python, JavaScript and Golang.
This section goes through each of the pillars and provides a concise description of our automation code and technologies covering the respective pillars.
Finally, we present an example workflow.
Even though the provided implementation of ExaC has been used for several of our experiments, it should be considered a minimal working example to demonstrate the concept and not a mature code base.
One of the added advantages of this modular design is the potential evolutive improvement. 
Accordingly, successful ExaC code bases and code components will be replicated, modified, and improved. 

\begin{figure*}[htpb!]
  \includegraphics[width=\textwidth]{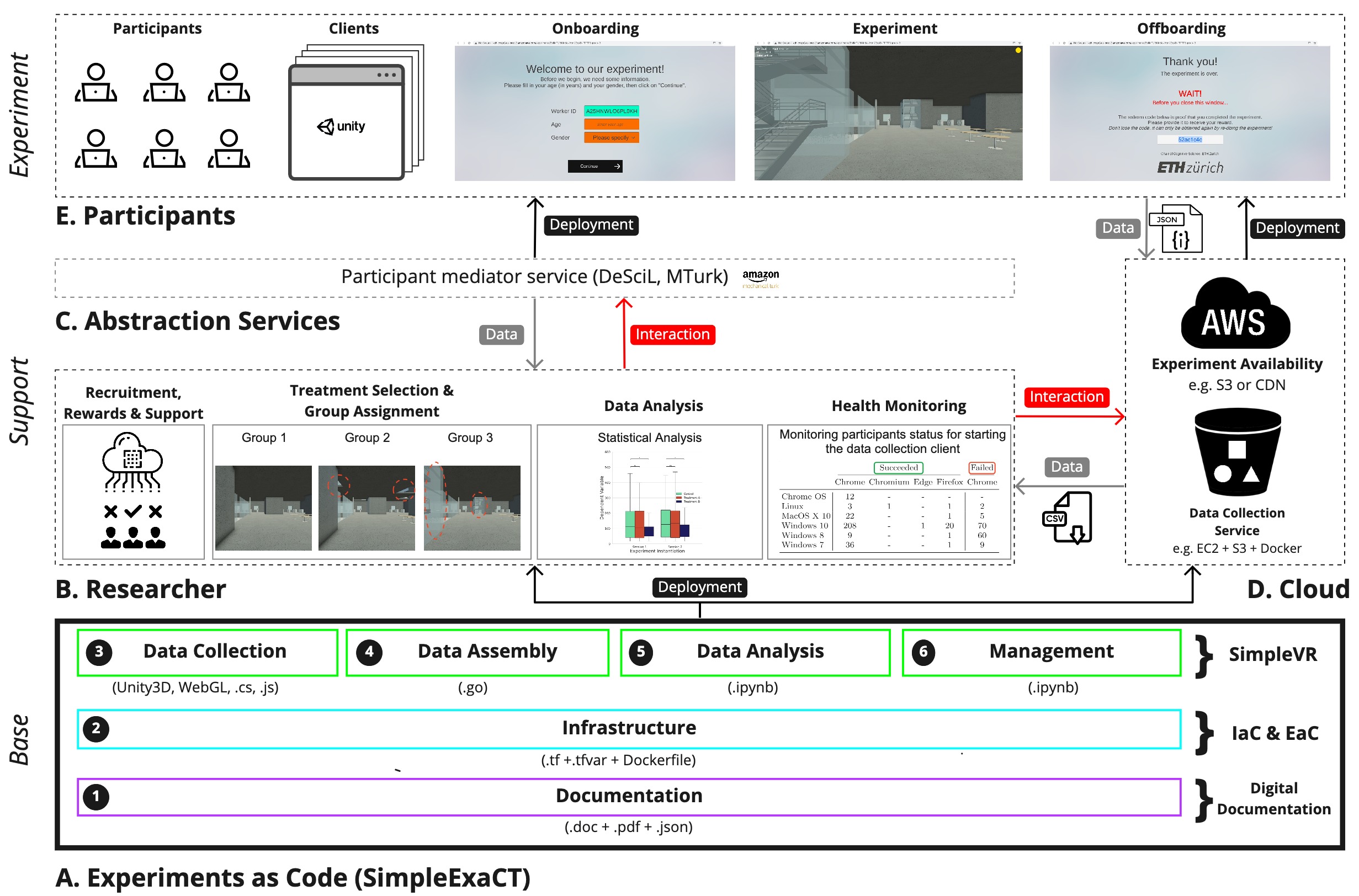}
\caption{Overview of the ExaC implementation and instantiation for the purpose of our case-study experiment (see Sec.~\ref{sec:case:study}). The illustration integrates the ExaC implementation (Base), the instantiated services and infrastructure (Support) and the data collection client (Experiment).} 
\label{fig:experiment:as:code}       
\end{figure*}

In our ExaC paradigm implementation, we distinguish between three layers (see Fig.~\ref{fig:experiment:as:code}, A).
The top layer consists of the software stack that describes what the experiment is (see Fig.~\ref{fig:experiment:as:code}, A3, A4, A5, \& A6).
The middle layer consists of the software stack that describes how the experiment is run (see Fig.~\ref{fig:experiment:as:code}, A2).
The bottom layer consists of the digital documentation management (see Fig.~\ref{fig:experiment:as:code}, A1).
The distinction between the stacks is similar to the differentiation between dev-stack and a devops-stack in software service development.
In the top layer, we provide an implementation of the four pillars with our proof-of-concept framework SimpleVR, analogous to the dev-stack.
In the middle layer, we provide software defining run-time environments, automation, provisioning, and deployment, analogous to the devops-stack.
Our ExaC implementation is the composite of these three layers, which we refer to as SimpleExaCT \citep{SimpleExaCTGithubRepo}.
The middle layer of the ExaC implementation supports Preproducibility by automating the deployment through cloud computing concepts, namely EaC and IaC.
We containerized the top layer services using EaC (e.g., a Docker image).
The middle layer provisions the infrastructure and deploy the containerized services using IaC (e.g., Terraform).

Since the system is meant to support reproducible experiments, we pay particular attention to the top layer that contains the experiment defining code.
To that end, we address its design considerations in light of our case study needs, specifically remote VR experimentation \citep{rivu2021remote}.
For the most part, conducting VR studies has been the privilege of well-funded institutions because of the prohibitive costs of large-scale experimental data collection.
We address this issue with a minimalistic ad-hoc VR framework, SimpleVR, by reducing the computational cost of large-scale data collection.
In traditional lab-based VR experiments, the researchers need to provide the computational power to run each of the VR experimental trials.
This is done through a central server or offloaded to the lab computers.

In contrast, SimpleVR offloads the VR computations to the participants' browser and requires the researchers to provide only the infrastructure needed to capture the transferred data. 
This is supported by new web browsers that provide local browser storage, service workers, access to 3D rendering capabilities, in-browser databases, and the ability to execute code with low overhead through web-assembly.
It needs to be noted that the purpose of this paper is not to present the development of a feature rich VR framework. 
Instead, we focus on the novelty of the ExaC paradigm, and this minimal framework (i.e., SimpleVR) was developed to bring the experiment to the browser and minimally cover the top layer in the 6 pillar taxonomy.

SimpleVR represents the top layer of the ExaC code Base (see Fig.~\ref{fig:experiment:as:code}, A3, A4, A5 \& A6).
When deployed, the instances of the four composable component groups serve a number of purposes.
The first component group (see Fig.~\ref{fig:experiment:as:code},  A3) defines the data collection software which is displayed in the client browser (see Fig.~\ref{fig:experiment:as:code}, E). Trials of independent human participants run within their web browser to collect data.
The second component group (see Fig.~\ref{fig:experiment:as:code}, A4) defines the data assembly service which is deployed in the cloud (see Fig.~\ref{fig:experiment:as:code}, D). 
This component group provides an API to the first component detailing how to collect and store the data.
The third component (see Fig.~\ref{fig:experiment:as:code}, A5) defines the data analysis service that allows to process data and gain insights (see Fig.~\ref{fig:experiment:as:code}, B).
This component group interacts with the data assembly.
The fourth  component group (see Fig.~\ref{fig:experiment:as:code}, A6) defines the data management service that allows the researcher to monitor and manage the experiment (see Fig.~\ref{fig:experiment:as:code}, B \& C) and interacts with all of the three previous component groups.
Once the data collection and data assembly services are ready, the experiment deployment infrastructure distributes the data collection software to the participants.

\subsection{ExaC implementation -- A}

The SimpleExaCT code base consists of the digital documentation, the IaC and EaC technologies and SimpleVR.
We refer to it as the \emph{Base} (see Fig.~\ref{fig:experiment:as:code}, A). 
A description of the implementation of the proof of concept is provided in Section 1 of the Supplementary Materials.

\subsection{Demonstrative Case Study}
\label{sec:case:study}

To demonstrate the power of the ExaC paradigm, we use the SimpleExaCT implementation in a desktop VR behavioral experiment in the domain of Cognitive Science.
The experiment consists in participants finding their way to different targets within a building. The experiment evaluates the effect of two architectural variations `Treatment A' and `Treatment B' on the wayfinding performance of participants in comparison to the unmodified building `Control'.
We selected this remote destkop VR experiment for our proof of concept as it doesn't require specialized equipment beyond a capable computer and a browser, and is robust under a wide range of uncontrolled variations, e.g., variations in screen sizes, or distance to the screen. This enabled us to easily showcase the use remote data collection and demonstrate this kind of vertical scaling.
This experiment was performed in April 2020 in three sessions, during the COVID-19 pandemic and lockdowns. The use of SimpleVR allowed us to bring the VR experiments to the browser and our SimpleExaCT implementation allowed us to communicate and document the procedure following the preproducbility concept.

The data collection in the original experiment was spread out across days to obtain enough participants according to the power analysis. 
Although for the original experiment, all sessions where aggregated in the data pool, in terms of independent instantiations, the three sessions can be understood as three independent experiments.
The third session did not collect data on all treatments because only some treatments required more data.
Hence, the results presented in this paper include only a comparison between the two full instantiations, sessions 1 and 2. 
For demonstration purposes and to avoid conflict with the detailed publications of the experiment, here we refer in general terms to: 1) the control and applied treatments which were architectural variations on a building, and 2) the dependent variable used to measure the participant's performance.
Details on the experiment and the behavioral analysis of the MTurk participant data can be found here \citep{Gath-Morad2021Spatial}. Details on the validation of an Agent-Based Model (ABM) using this data can be found here \citep{Gath-Morad2020Towards}.

Participants were recruited online through MTurk Human Intelligence Tasks (HITs).
The participants were filtered according to the external requirements of informed consent, technical ability to display the experiment and completion of the experiment.
We only used data from the participants whose web-browser was deemed capable of displaying the experiment, who agreed on the consent screen, and who completed the task with the correct response to the challenge. 
149 participants fulfilled these criteria. Participants' age ranged between 18 years to 59 years.
The mean age was 33.7 years with a Standard Deviation of 6.8 years.
The methods and experiment described in this section were performed in accordance with the relevant guidelines and regulations following the approval of the study by ETH Zurich's Research Ethics Committee. 
Two inclusion criteria for the study were English proficiency, corrected-to-normal or normal vision. Color blind individuals were excluded from participating in the study.
All participants signed an informed consent form. 
The average time taken to complete the experiment was 20 minutes and the average monetary compensation was 4.5 USD. If participants completed the experiment in less than 20 minutes they received 1 USD bonus. The average compensation was 5.4 USD.

\paragraph{Experimental Protocol}
The protocol of the experiment can be found in the SimpleExaCT repository, \citep{SimpleExaCTGithubRepo} and a complete description of the demonstrative experiment content, the experimental protocol, materials, and data analysis can be found at \cite{Gath-Morad2020Towards}. 
Overall, the experiment tested how changes in the configuration of buildings may affect human navigation performance. 

Figure \ref{fig:experiment:as:code} shows how ExaC was used to deploy the experiment. 
An abstraction layer (Figure\ref{fig:experiment:as:code}, C) was in interfacing with the participant procurement service.
For logistical reasons, we opted for the DeSciL and MTurk as specific service providers in charge of communication with both the researcher and with the study participants.
Nonetheless, we provide scripts to interface with MTurk directly within the SimpleExaCT implementation.
Communication with participants mainly consisted of deploying the HITs for them to perform the experiments.

Participants had a a first person perspective of the virtual scene in each building. Participants' origin was set to the same location across the three building groups (see Fig.~\ref{fig:experiment:as:code}, E).
Navigation was performed using keyboard and mouse.
Participants had to digitally sign a consent form and complete the onboarding before the experimental trials and the offboarding afterwards.
The onboarding ensured the correct assignment to one of the three building treatments and the offboarding the appropriate reward.
The experiment itself mainly consisted of two parts, a training phase and 6 navigation trials.
The mean duration of the training phase 5 minutes. During training,  participants learned how to move in the virtual environment using a mouse and keyboard. 
This is a common procedure to ensure that participants are familiar with the controls to reduce adverse effects \citep{thrash2015evaluation,grubel2017evaluation}.
Upon successful completion of the practice scene, participants had to complete 6 trials in which they were asked to find a specific destination inside the building.

The participants’ positions and orientations were recor\-ded every frame (roughly every 0.02 seconds) using a cloud infrastructure to deploy and retrieve participants’ data (see Fig.~\ref{fig:experiment:as:code}, C \& D).
The research team (see Fig.~\ref{fig:experiment:as:code}, B) was able to manage and monitor participants, hardware and software status from their computers.
The abstraction services provided interactions with the management of participants (see Fig.~\ref{fig:experiment:as:code}, B \& C).
The cloud infrastructure provided interactions with both the deployment of the experiment content and access to the collected data (see Fig.\ref{fig:experiment:as:code}, B \& D).
Using the analysis scripts, the research team was able to visualize and analyze the data collected immediately after the experiment was concluded (see Fig.~\ref{fig:experiment:as:code}, B). 

\subsection{Results}

\begin{figure}[htpb!]
  \includegraphics[width=\columnwidth]{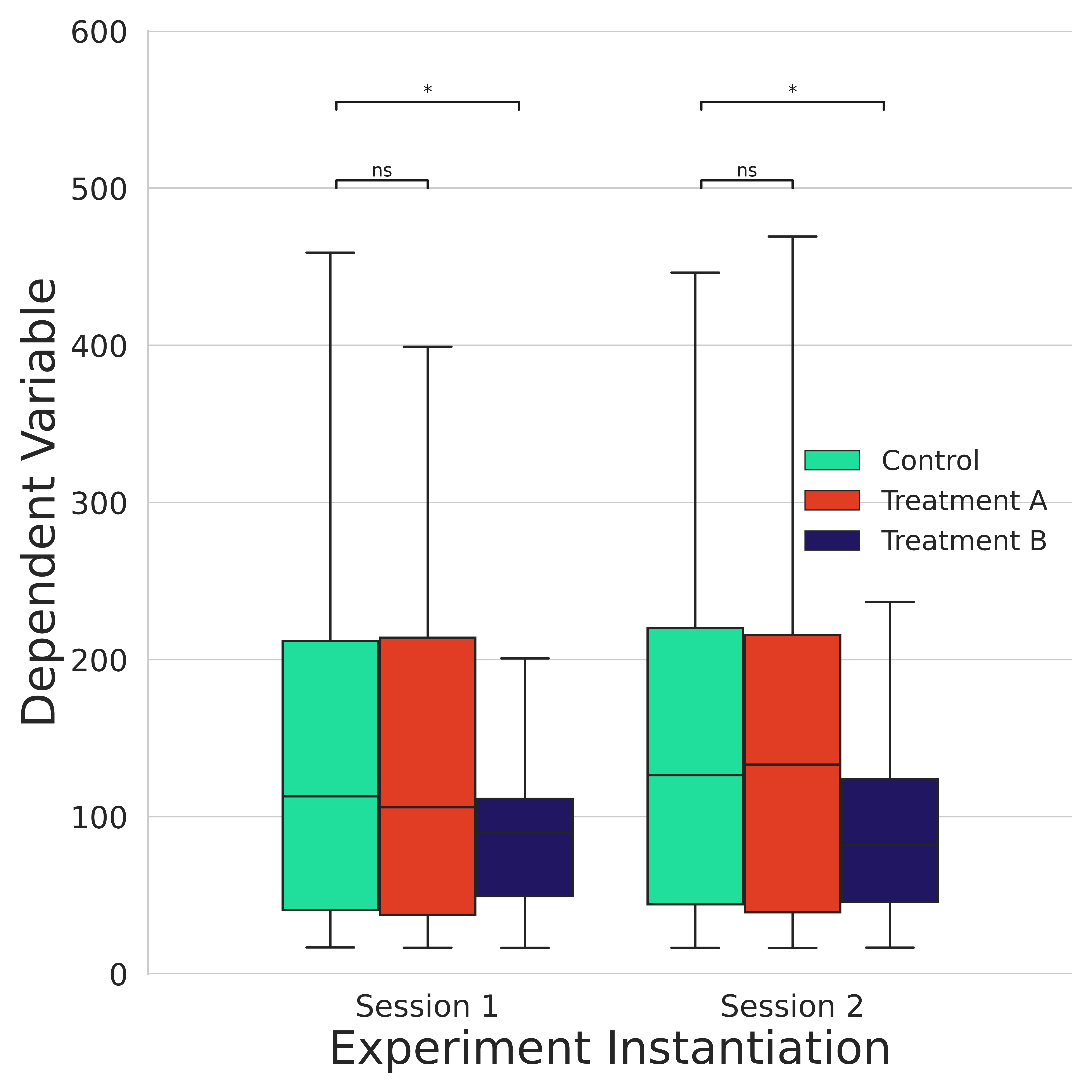}
\caption{The same pattern of results is shown by the two complete instantiations of the experiment, lower value on the dependent variable means and improvement in performance. Session 1 (n=275), Session 2 (n=495). For an in-depth data analysis of the experimental results see \citep{Gath-Morad2021Spatial}.} 
\label{fig:experiment:reproducibility}       
\end{figure}

\sloppy \paragraph{Reproducibility} is assessed by independently evaluating the results of the two complete instantiations with the Linear Mixed Effect Model, $\mbox{DependentVariable} \sim \mbox{Treatment}  + ( 1 | \mbox{Participant})$.
Figure \ref{fig:experiment:reproducibility} shows that in the complete instantiations of the experiment, the same pattern of results can be observed, treatment A has no significant effect, while treatment B provided a significant ($p<0.05$) performance improvement. 

\paragraph{Auditability, Debuggability, \& Reusability} are shown through a simulated third party auditing the experiment and describing bugs found while using the code. This is documented in a GitHub issue `Example 1 - Bug fix'. As a response to this GitHub issue, the maintainer identifies the cause of the bug and creates a new version reusing most of the code.  Additionally, the GitHub issue `Example 2 - Reuse', shows a simulated interaction with a researcher reusing SimpleExaCT for an online version of another desktop VR wayfinding experiment \citep{Gath-Morad2020how}. Here, the researcher adds the experiment assets, i.e., the building and building variations, and implements the experiment mechanics, e.g., door frames changing in color. These modifications are limited to the data collection module, while all the other modules are reused without changes. All Github issues presented here as examples are linked with their respective code. Interested third parties are encouraged to check the repository Issues section in addition to auditing and forking the repository.

\paragraph{Scalability}

We investigate the access to the experiment beyond the 149 participants that successfully took part in the experiment by reviewing the raw accesses to the experiment.
This includes all the interactions with the experimental data collection (i.e., MTurkers that clicked on the experiment link, see Fig.~\ref{fig:experiment:as:code}, E).
In total, 462 participants accessed the experiment, 316 (68\%) had computers that were deemed capable of displaying the experiment, while the remaining 146 (32\%) were not.
Of the participants starting the experiment, 149 (47\%) participants also completed the experiment.
The participants' inferred Operating System and browsers are reported in Table \ref{tab:user:agent}.

\begin{table}[htbp!]
    \centering
    \resizebox{\columnwidth}{!}{%
\begin{tabular}{l|ccc|c}
            \toprule
            \multicolumn{1}{c}{} & \multicolumn{3}{c}{Succeeded} & \multicolumn{1}{c}{Failed} \\
            \cmidrule(rl){2-4} \cmidrule(rl){5-5}
      \multicolumn{1}{c}{} & Chrome &  Firefox & \multicolumn{1}{c}{Other} & \multicolumn{1}{c}{Chrome} \\
\midrule
Chrome OS   & 12  & -  & - & - \\
Linux       & 3   & 1  & 1 & 2 \\
MacOS X 10  & 22  & 1  & - & 5 \\
Windows 10  & 208 & 20 & 1 & 70\\
Windows 8   & 9   & 1  & - & 60 \\ 
Windows 7   & 36  & 1  & - & 9 \\
\midrule
Total       & 290  & 24  & 2 & 146 \\
\bottomrule
\end{tabular}
}
    \caption{Excerpt from the participant monitoring. Operating system and browser are inferred from the self reported browser user agents. The client was successfully rendered on 316 computers and failed to render on 146 computers. }
    \label{tab:user:agent}
\end{table}

\section{Discussion}

Behavioral experiments in VR often require time consuming setup and diverse expertise in specialized hardware (e.g., physiological sensors \citep{weibel2018virtual,hackman2019neighborhood}) or infrastructure well beyond the capacity of smaller research groups and researchers \citep{fischer2020investigating,spielhofer2020physiological}.
This difficulty, coupled with the implicit information required to reproduce an experiment, has contributed to the reproducibility crisis across many disciplines.
We believe that this crisis can be mitigated through the use of emergent technologies such as the ones behind the ``\underline{\hspace{0.5cm}} as Code'' and ``\underline{\hspace{0.5cm}} as a Service'' movements.
The ``\underline{\hspace{0.5cm}} as Code'' technologies provide a tool-set for automation and reproducible software environments.
Together with VR frameworks, this enables the implementation of the so far conceptual notion of `Preproducibility'.
The ``\underline{\hspace{0.5cm}} as a Service'' technologies provide outsourcing and abstraction layers that can be used to interface with real world needs such as the recruiting and rewarding of participants. 


To explore the applicability of ``\underline{\hspace{0.5cm}} as Code'' and ``\underline{\hspace{0.5cm}} as a Service'' to behavioral experiments in VR, we first define a taxonomy for behavioral studies in VR (see Fig.~\ref{fig:six:pillars}) to define a scope in which automation could and should occur.
We argue that there is a need for such a taxonomy to create a common framework for shareable work and modularity that can enable the reuse of individual components. 
Second, we proposed a definition of the ExaC paradigm as providing declarative or imperative automation code and digital documentation that corresponds to each of the experiment taxonomy pillars (see Fig.~\ref{fig:experiment:as:code:abstract}).
By publishing their codebase, researchers offer concrete means of reproducing their research.
Hence, enabling their experiments to be reproduced, audited, debugged, reused, and scaled by other researchers.
Beyond replication, their research can then be systematically varied to uncover robustness and generalizability.
Additionally, different experimental approaches could be adapted to the same experiment, thus enabling triangulation \citep{munafo2018robust}.
In triangulation, the same concept is tested with different approaches to ensure generalizability.

%
To elucidate the applicability of the ExaC paradigm to support Preproduciblity in VR experiments, we present a proof-of-concept implementation of the ExaC and apply it in a VR case-study in the domain of cognitive science.
The proof-of-concept implementation of ExaC consists of the digital documentation, the infrastructure \& environment definitions, and the SimpleVR framework.

Our proof-of-concept implementation of ExaC requires a plethora of technologies.
This might seem counterproductive at first sight as it could require strong expertise to master.
The technologies and software stacks range from Docker, Terraform, Jupyter-Lab, Unity, to WebAssembly, and WebGL.
On top, they require using several programming languages, Golang, Python, JavaScript, and C\#.
To add to this complexity, it has to integrate services that serve as an abstract interface to the physical world, such as DeSciL and MTurk.
However, they are some of the best technologies available for each specific task and there is currently no one-size-fits-all solution. 
The additional effort will be amortized by reusing ExaC code bases.
Even though monolithic highly coupled software can provide a more homogeneous user experience, we believe it would be difficult for it to keep up with the rapid advancement of the technology.
Furthermore, these loosely coupled modular components can be exchanged as technologies progress or as requirements change.

Beyond the fast-paced development of technology, the loose coupling actually reduces end-user complexity.
End users are only required to change the components they work on, enabling them to consider all other components \emph{black boxes}.
In the context of VR, this means that using ExaC allows experimenters to focus on experiment content rather than any aspect of deployment and provisioning.
Through loose coupling and automation we can tame the complexity and implement for the first time the concept of Preproducibility.

There remain many technical challenges to be tackled for more general ExaC implementations beyond the proof of concept. For example, interfaces for specialized local or remote hardware, and specific protocols for composition, description and management of services in the context of experiments need to be developed.

In our case study, we leveraged the ExaC paradigm to scale and communicate the experimental efforts.
This enabled us to overcome the limitations of the COVID-19 pandemic lockdown that put on hold in-person experiments.
The challenges regarding reproducibility are specific to the experiment's nature, requirements and constraints. We chose this simple desktop VR experiment as proof-of-concept to showcase all the benefits of its as code representation.
We showcase the reproducibility of our proof-of-concept experiment by analyzing two complete instantiations of the experiment.
We show that the pattern of results remains the same, i.e., `Treatment A' displays a null effect while `Treatment B' displays a statistically significant improvement in comparison to the `Control' building. 
We showcase the auditability, debuggability, and reusability of the experiment by performing a task common in open source projects, i.e., a simulated auditor opened a GitHub issue and provided a description of a bug and a fix request, while the repository maintainer identified the required changes in the code and developed that fix. Additionally, we show the implementation of a second experiment by changing the data collection module and reusing the rest of the components.
We demonstrate the experiment scaling in two ways, by enabling several sessions of data collection (horizontal scaling) and by recruiting participants from a big remote participant pool in comparison to the usually smaller local participant pools (vertical scaling).
Even though our VR experiment was graphically demanding and data heavy, we reached 462 participants of whom the majority (68\%) had computers that were deemed capable of running the desktop VR experiment.
From this pool, 149 participants successfully completed our experiment.
This showcases the scaling potential of this experiment.

Many experiments pose more challenging reproducibility requirements, e.g., interfacing with specialized local hardware, or behavioural response changes due to nuanced variations in the setup, sensors or equipment.
Researchers can document as ExaC the solutions required to successfully run the experiment, e.g., local server setup and hardware interfaces, sensor configuration, or specialized hardware setup. This would enable themselves and third parties to reuse the solutions and would make the nuances transparent.
With an ExaC ecosystem, researchers could focus on the specifics of their experiments while reusing parts of existing and tested experimental setups.

\section{Conclusion}

Communicating scientific results requires enumerating, recording, and reporting those things that ``cannot with advantage be omitted'' \citep[p.44]{popper1992open}. 
Depending on the discipline, Preproducibility \citep{stark2018before} might require information about materials (including organisms and their care), instruments, and procedures; experimental design; raw data at the instrument level; algorithms used to process the raw data; computational tools used in analyses, including any parameter settings or ad-hoc choices; code, processed data and software build environments; or even analyses that were tried and abandoned.
In this paper we provide a conceptual grounding for ExaC based on the Preproducibility.
We provide a taxonomy of the components for a practical implementation and provide a proof-of-concept implementation showcasing its reproducibility, auditability, debuggability, reusability, and scalability of the ExaC paradigm.

There are other systems envisioned to define precise courses of action, including what is known as \emph{Smart Contracts} a fundamental building block of \emph{decentralized apps} \citep{wang2019blockchain}, based on blockchain technology. 
However, in contrast to what is attempted in the area of \emph{Social Computing} \citep{messinger2009virtual}, we are not primarily interested in automation but in reproducibility.
A desirable positive reward-based system could take advantage of both technologies.
Stake holders could create rewards pools that encourage the reproduction of experiments in a decentralized manner.
Interested parties could earn these rewards by providing proof of reproducibility or reproducing experiments.
All this could be scaffolded on \emph{decentralized apps}, blockchain technology and ExaC.

It is well-known that automation of social processes can also have undesirable implications \citep{helbing2015automation}.
These problems range from loss of creative freedoms to new forms of exploitative, or alienating labor.
We would like to distance ourselves from these practices.
We believe that ExaC based experiments can mitigate these problems by ensuring transparent and auditable interactions with a society where fair practices such as rewards based on decent wages are enforced.

The ExaC paradigm is applicable beyond the narrow frame of VR experiments if we can define consistent abstraction interfaces to non-virtual components.
Going beyond outsourcing research to specialized groups and facilities, the world advances with services that provide additional layers of abstraction.
Such services allow us to interface with non-virtual components (e.g., MTurk, DeSciL).
Moreover, the ability to interface with reality from virtual worlds would enable innovative experiments based on virtual components, e.g., digital twins \citep{grieves2017digital,grubel2021fused}, enhanced ABMs, and multi-agent-systems \citep{helbing2012agent,grubel2019cognitive,schaumann2016computational,aguilar2014scalable,aguilar2015mixed,aguilar2016automatic,aguilar2019mass,Gath-Morad2020Towards,zhao2020assessing,zhao2022fire} and medical applications (\citeauthor{brander2021virtual}, \citeyear{brander2021virtual}).
In combination with the ExaC paradigm, it becomes possible to define reproducible virtual(ized) and mixed-virtual experiments in the real world.

There is a great potential in extending the ecosystem of ``\underline{\hspace{0.5cm}} as Code'' and ``\underline{\hspace{0.5cm}} as a Service'' to research facilities.
Complex experiments in physics, chemistry, biology, and engineering could be transformed to a specification within the ExaC paradigm.
This specification could be automatically sent to specialized facilities possessing unique equipment and domain knowledge of the best practices, ethical requirements, and efficient procedures.
However, there are clear limitations as to how far this interfacing with non-virtual components can reach.
For example, there could be sensitive interventions that could be dangerous or unethical to automate.
Nonetheless, when the ExaC paradigm can be applied, it contributes to the democratization and improvement of the quality of research.

The ExaC paradigm and implementation could also provide an answer to challenging questions about reproducibility.
For example, the inability to reproduce certain results in behavioral science could potentially be due to behavioral drifts.
The individuals participating in a study could represent a group that has gradually changed its behavioral patterns due to societal, technological, environmental, ecological, or economical changes.
When a study is reproduced, the previous results no longer hold as the group has shifted from previous behavioral patterns.
By providing automation code, it would be possible for stakeholders to easily and continuously reproduce experiments and consequently observe the behavioral drifts over longer periods and across participants' compositions. 

The ExaC paradigm can also provide answers beyond behavioral drift by improving the generalizability of research results.
Reusing flawed experiments over and over again could encourage the prevalence and perpetuation of bad practices.
To counter this, we believe the ExaC ecosystem would need to adopt, document, and introduce best practices.
However, the evolutive aspect of ExaC, if widely adopted, should also on its own weed out bad experiments and replace them with better versions.
Additionally and beyond replication, triangulation \citep{munafo2018robust} can be applied to systematically vary experiments to uncover the underlying fundamental principles.
Given the ability to easily reuse existing experiments, triangulated versions can be produced to shed light on phenomena from a different perspective.

In the future, we plan to further develop our proof of concept, providing more features to better achieve reproducibility.
Previous work on VR has already produced robust frameworks for data collection (e.g., EVE \citep{grubel2016eve}, \citep{zhao2018networked}).
We would like to integrate these frameworks to provide improved participant management but also tools to aid the quick and effective design of experiments and multi-user environments \citep{zhao2018networked}.
Additionally, we would like to improve the analysis and data management of experiments by integrating complete machine learning lifecycles \citep{aguilar2021ease}.
Our current implementation of ExaC, SimpleExaCT, limited itself to a descriptive non-automated digital documentation.
However, the digital documentation still holds a lot of potential for automation and enhancements.
For instance, the pre-registration of a study (e.g. with the Open Science Framework; OSF) could be automated.
DOIs for the composable components can be minted using the API of services such as CERN's Zenodo making the experiment easily available to other researchers.
The file \texttt{services.json} could serve as an information source to procure and manage the required external abstraction services including opening accounts, procuring resources, and exchanging information. 
Finally, we want to extend the proposed taxonomy to more complete ontological relations \citep{noy1998representing, soldatova2006ontology} and follow best practices in their design \citep{soldatova2005current}.
Combining all these ideas could enable the creation of robust knowledge databases containing ExaC experiments and results.

In summary, we believe that the ExaC paradigm has a great potential to enhance cooperation, enable equitable access to research resources, underpin research in VR, especially in times of limited physical interaction.
Moreover, we built a proof of concept based on the proposed taxonomy and demonstrate the possibility to implement Preproducibility in a scalable online web-based VR behavioral experiment. Through this proof of concept ExaC implementation we showcase reproducibility, auditability, debuggability, reusability and scalability.
The adoption of the ExaC paradigm can improve the quality of research across several domains, by providing reproducibility, auditability, debuggability, reusability, and scalability to experimental research.

\section{Acknowledgements}
We would like to thank Giordano Giannoccolo and Joris Stemmle from DeSciL for their support in the data collection and MTurk experiment deployment experience.

\subsection{Funding}

This research has been funded by the Chair of Cognitive Science and the Data Science, Services and Systems Laboratory at ETH Z\"urich.

\subsection{Conflicts of interest/Competing interests}
The authors declare that they have no conflict of interest.

\subsection{Availability of data and material (data transparency)}

For the interested reader, a constrained reproducible demonstration is provided.
This constrained reproducible demonstration diverges from the case study as it uses free/test services where possible, and is fully automated directly through MTurk instead of using DeSciL. 
To reproduce the constrained demonstrative experiment, an account in Amazon's Mechanical Turk and Amazon Web Services are required.
The total cost of reproducing the full experiment is estimated below 800 USD (with an average payout of 5.4 USD per participant and about 140 participants) as of January, 2021.

\subsection{Code Availability}
The workflow required to reproduce the experiment is quite simple.
The instructions are provided in the \texttt{READ\-ME} of the github repository \citep{SimpleExaCTGithubRepo}. The checklist is summarized here:

\begin{enumerate}
	\item Fill in or customize the variables in the Terraform script to access cloud services (e.g., AWS, MTurk etc).
	\item Provision and deploy experiment infrastructure: \\ \shellcmd{terraform init}, and \shellcmd{terraform apply}
	\item Release experiment trials (HITs) for participants to complete through the provided MTurk's scripts
	\item Manage and Monitor the experiment through the management scripts.
	\item Download your data and analysis results.
\end{enumerate}

\subsection{Authors' contributions}

Contributions are organized by CRediT (Contributor Roles Taxonomy) and authors are listed in alphabetical order of the surname.

\noindent \emph{Conceptualization}: L. Aguilar, M. Gath-Morad, J. Gr\"u\-bel, D. Helbing, C. H\"olscher, R. W. Sumner, C. Zhang

\noindent \emph{Data curation}: L. Aguilar

\noindent \emph{Formal Analysis}: L. Aguilar

\noindent \emph{Funding acquisition}: C. H\"olscher, C. Zhang

\noindent \emph{Investigation}: L. Aguilar, M. Gath-Morad, S. Wehrli

\noindent \emph{Methodology}: L. Aguilar, M. Gath-Morad, J. Gr\"u\-bel

\noindent \emph{Project administration}: C. H\"olscher

\noindent \emph{Resources}: C. H\"olscher, S. Wehrli, C. Zhang

\noindent \emph{Software}: L. Aguilar, J. Ermatinger, S. Wehrli, H. Zhao

\noindent \emph{Supervision}: C. H\"olscher, R. W. Sumner, C. Zhang

\noindent \emph{Validation}: L. Aguilar

\noindent \emph{Visualization}: L. Aguilar, M. Gath-Morad, J. Gr\"u\-bel

\noindent \emph{Writing – original draft}: L. Aguilar, M. Gath-Morad, J. Gr\"ubel

\noindent \emph{Writing – review \& editing}: L. Aguilar, M. Gath-Morad, J. Gr\"ubel, D. Helbing, C. H\"olscher, R. W. Sumner, S. Wehrli, H. Zhao, C. Zhang


\subsection{Compliance with Ethical Standards}

The research with human participants was approved by the Research Ethics Committee of ETH Z\"urich (2020-N-24).
The participants were informed on the study goal and gave informed consent and accepted the publication of appropriately anonymized data.

\bibliographystyle{spbasic}  
\bibliography{ref}

\end{document}


\title{Supplementary Information: Experiments as Code \\
\large A Concept for Reproducible, Auditable, Debuggable, Reusable, \& Scalable Experiments
}

\author{
Leonel Aguilar$^{1,2,\dagger}$ \href{https://orcid.org/0000-0001-6864-4492}{\includegraphics[scale=0.09]{figures/orcid.png}}\and
Michal Gath-Morad$^{1,\dagger}$ \href{https://orcid.org/0000-0001-7673-6290}{\includegraphics[scale=0.09]{figures/orcid.png}}\and	
Jascha Gr\"ubel$^{1,3,\dagger}$  \href{https://orcid.org/0000-0002-6428-4685}{\includegraphics[scale=0.09]{figures/orcid.png}}\and
Jasper Ermatinger$^1$ \and
Hantao Zhao$^{1}$ \href{https://orcid.org/0000-0003-0398-3842}{\includegraphics[scale=0.09]{figures/orcid.png}}\and
Stefan Wehrli$^5$ \href{https://orcid.org/0000-0001-9845-9028)}{\includegraphics[scale=0.09]{figures/orcid.png}}\and
Robert W. Sumner$^3$ \and
Ce Zhang$^{2}$\and
Dirk Helbing$^5,^6$ \and
Christoph H\"olscher$^{1,5}$\href{https://orcid.org/0000-0002-5536-6582}{\includegraphics[scale=0.09]{figures/orcid.png}}
}

\maketitle

\renewcommand{\theequation}{S\arabic{equation}}
\renewcommand{\thefigure}{S\arabic{figure}}
\renewcommand{\bibnumfmt}[1]{[S#1]}
\renewcommand{\citenumfont}[1]{S#1}

Contact: Leonel Aguilar\\ leonel.aguilar@gess.ethz.ch\\
   $^1$ Chair of Cognitive Science, ETH Z\"urich\\
   $^2$ Data Science, Systems and Services Group, ETH Z\"urich\\
   $^3$ Game Technology Center, ETH Z\"urich\\
   $^5$ Decision Science Laboratory, ETH Z\"urich\\
   $^6$ Chair of Computational Social Science, ETH Z\"rich\\
   $\dagger$ These authors provided equivalent contributions
\twocolumn
\section{Proof of Concept ExaC implementation - A}

This section provides details on the components of ExaC implementation of a simple wayfinding experiment in the field of Cognitive Science.
All components are provided in the GitHub repository: \url{https://github.com/leaguilar/SimpleExaCT}. To handle the synchrony between the components' versions we used the `mono repo' approach, where every component is contained in a subfolder of the same repository. In other words, this single repository holds all the experiment's code, e.g., documentation, provision, deployment, data collection, data assembly, data analysis, and experiment management code. Additionally, every component uses its specific versioning pinning mechanism, e.g., terraform's version pinning, docker image definition.

\subsection{Documentation -- A1}

The digital documentation for this example implementation consists of the necessary meta-information.
It includes traditional software metadata, the \texttt{README}, \texttt{AU\-THORS}, and \texttt{LICENSE} files along with the experiment specific documentation such as the experiment protocol. 
Incepting a good protocol is still more of an art but certain guidelines are available such as the VR-Check \citep{krohn2020multidimensional}.
While it focuses on developing neuroscience training in VR, it also offers an overview of key attributes to consider when drafting any experiment protocol (see the Supplementary Information Section 3).
Apart from the traditional software requirement documentation (e.g., \texttt{package.json}, \texttt{requirements.txt}, and \texttt{Pip\-file}) and the traditional experiment documentation (e.g., the protocol) we include a service documentation file  \texttt{servi\-ces.json}, denoting the requirements in terms of external services (i.e., abstraction layer services, and infrastructure services) required for the experiment (e.g., MTurk, AWS).

\subsection{Infrastructure -- A2}

SimpleExaCT defines the required run-time environments with Docker (i.e. EaC; see Fig.~\ref{fig:experiment:as:code}, A2) and automates the provisioning and deployment of the required experimental infrastructure through \emph{Terraform} scripts (i.e. IaC; see Fig.~\ref{fig:experiment:as:code}, A2).
The script creates an S3 bucket and uploads the static content, reserves the computing infrastructure for the data assembly service, and deploys it.
All this is initiated directly from the researchers' computer.
Once the data collection has finished, the \emph{Terraform} scripts automate the stop and release of all cloud resources used in the experiment.
 
\subsection{Data collection -- A3}
The participants join the experiment with their web browser.
In practice, a link is provided and the experimental data collection client is loaded as any other web-page.
As participants visit this link, the experiment goes through 3 phases: the \emph{Onboarding} of participants, the participants' engagement in the \emph{Experiment} trial and finally their \emph{Offboarding} (see Fig.~\ref{fig:experiment:as:code}, E). 
SimpleVR contains scripts and templates that instrument this data collection procedure (see Fig.~\ref{fig:experiment:as:code}, A3).

In the onboarding, we perform two checks to ensure that both Unity can be started and that it runs at a reasonable frame rate.
We use Unity's support of web-assembly with WebGL as ``build'' target to offload the VR computations to the web-browser.
To determine if the web-assembly module can be loaded, the browser compatibility and rendering capabilities are detected through Unity.
The prospective quality of rendering is tested using the test suite of mapboxgl (a JavaScript library).
If the requirements are fulfilled, the experiment content is loaded.
The Unity-based data collection client receives the parameters to instantiate the experiments' content along with authentication for participants' payment.

In the experiment, the specific treatment tested in this trial is instantiated for data collection, and the content is rendered in the participants' browser (see Fig.~\ref{fig:experiment:as:code}, E).
The whole experimental trial is distributed as static content.
For example, in the presented case study through an AWS S3 Bucket.
We used Unity to render, control, record, and transfer movement data from the virtual environment on the participants' computer.
The collected experimental data is sent back to the data assembly service continuously throughout the trial.
The client can send different types of messages to the data assembly service to assemble, store, and make it available to the researcher.
For the simplicity of the demonstration, we implemented two main types, which are event messages and trajectory messages.
Event messages provide metadata about a specific event (e.g., agreeing on the consent form, completing a specific task).
Trajectories are recorded by logging participants' positions and orientations with varying time resolutions.
Trajectories represent heavier data transfers and thus are segmented into packages.
Data is transferred in the background while the participants complete the tasks (i.e., in 4.3Kb chunks) to the data collection service and stored in a secure cloud storage.

In the offboarding, once the participants have completed the trial, a response code is calculated by the data collection client to signal the conclusion of the experiment.
The code is based on an implementation of Salted Challenge Response Authentication Mechanism (SCRAM) \citep{newman2010salted}.
Here, a challenge code is generated and combined with a researcher-provided authentication called ``salt''.
This delivers the signal to external services like MTurk that participants have completed their task and that these services can transfer the payment. 

\subsection{Data Assembly -- A4}
The Data Assembly service provides the backend for data collection and securely stores and makes the data available to the researcher.
This service (see Fig.~\ref{fig:experiment:as:code}, A4) has been built as a lightweight go program due to the language's embedded concurrency features.
To deploy the Data Assembly service, we containerize it using Docker.
Heavy traffic experiments may require the scaling of the data collection service and running it as part of a Kubernetes cluster.
This enables having detailed control over the load balancer and reverse proxies for load sharing.

For the data transmission, the service receives the POST messages of the clients and stores them either in the service's local persistent storage.
The transmitted messages provide header and tail identifier messages.
The tails providing a checksum over the transferred data to ensure the correct reconstruction of the transferred data once all packages/chunks have been received.
Once the tail packet is received, the data for a specific trial is reconstructed and uploaded to a secure cloud storage (in the case study we use an AWS S3 bucket).
The data is transferred as JSONs for its reconstruction. The reconstructed data is assembled into CSV files and uploaded into a cloud storage solution (e.g., AWS S3 bucket).

The data transmission between the data collection service and the data assembly service is secured to prevent mishandling. 
It takes advantage of HTTPS (HTTP over TLS) enabling the encrypted transport of the data.
It has to be noted that modern web browsers will refuse to establish a connection over HTTPS with self-made SSL certificates.
If no SSL certificates are available, researchers will be forced to revert back to insecure transmission and protect the data transfers themselves, i.e. encrypt the data instead of the channel.

\subsection{Data Analysis -- A5}

To exemplify the data analysis, a Python Jupyter notebook with a minimal working example for our case study's analysis is provided. 
During the deployment of SimpleExaCT, a container with Jupyter-Lab on the local researcher computer is run to provide an interface to the bundled notebooks (see Fig.~\ref{fig:experiment:as:code}, A5 \& B ).
This component is well-qualified to serve as the basis for pre-registration for an analysis plan.

\subsection{Management -- A6}
The experiment is scaffolded with cloud infrastructure. 
It can be served to anyone with a capable web browser through a link to the static content and the corresponding \emph{Experiment} (see Fig.~\ref{fig:experiment:as:code}, E).
To address the difficulty of recruiting and securely rewarding the participants, we deployed the experiment through Amazon Mechanical Turk's (MTurk).

To monitor and manage both the hardware and participants\footnote{It has to be noted that monitoring and management should not be interpreted as \emph{surveillance}. It is a privacy-preserving, consensual, and non-intrusive check of the state of the experiment and health of the hardware.}, we include two further Python Jupyter notebooks, (see Fig.~\ref{fig:experiment:as:code}, A6).
We reuse the same dockerized local Jupyter-Lab instance mentioned in the data analysis to provide access to these notebooks (see Fig.~\ref{fig:experiment:as:code}, B). 
The participant monitoring and management notebooks interact with Amazon Mechanical Turk's API and serve to link our experiment with specific challenge codes and the correct treatment assignment, once the experiment is completed.
Additionally, it verifies the correctness of the responses to the challenges and rewards the participants. 
Through the hardware management notebook we monitor the state of the cloud infrastructure.
The current proof-of-concept implementation triggers an alarm sound when the Data Assembly service is unreachable.

\section{Experiments as Code in VR}
\label{app:ExaC:inVR}
This section provides an overview of the documented efforts to address components (e.g., tasks, procedures, information, and documentation) in the proposed six pillars taxonomy (see Fig.~\ref{fig:six:pillars}) of Experiments as Code.

\subsection{Documentation}
Experiments usually gather and require information that falls under the category \emph{documentation}.
This information enables the researcher to communicate aspects such as the experiment design, software usage procedures, data management plans, or tasks to be performed.
The documentation sets out the details of how to setup and run the experiment.
Additionally, it ensures comprehension within a research group and allows to validate a pilot study by comparing the actual outcome with the planned outcome.
Moreover, the documentation provides the necessary details to understand the data and produce analyses as well as writing papers.
Often the documentation is informal and only used for internal purposes such as communicating the experiment between co-authors and assistants implementing the experiment, but it encapsulates valuable information, especially if the experiment was to be modified.

To address these issues, pre-registration (reviewed and unreviewed) and analysis plans have been put forth to formalize experimental protocols \citep{van2016pre}, but so far these are only tentatively used in VR studies.
Such a protocol should be created first and iterated through revisions to ensure an experiment's effectiveness before it is implemented.
An experiment protocol must detail every step of an experiment from all sides.
It is important to report both from the participants perspective as well as from the process on the experimenters' side \citep{weibel2018virtual}.
This ensures that there is no step missing that may be seem obvious from one side but not the other.
Conducting pilot studies and rapid prototyping are crucial tools to ensure the quality of a protocol.
None of the reviewed frameworks explicitly address protocols.
However, many frameworks use configuration files and experimental flow descriptions that could be used to generate some form of automatic protocol.

\subsection{Infrastructure}
Even though the \emph{Infrastructure} is crucial to conducting an experiment, it is usually informally provisioned and deployed on a per case basis.
Reports usually only state the operating system, the game engine and sometimes the hardware specifications for data collection.
The issue is even starker because a large part of the infrastructure remains unspecified including the data processing, content delivery system, analysis equipment, and researcher equipment.
Only unusual configurations produce more detailed documentation such as multi-user VR \citep{moussaid2018virtual,zhao2018networked} but do not provide automation in the form of provisioning or deployment instructions.
The rise of IaC offers an opportunity to remove a major hurdle to equitably running VR experiments that to our knowledge so far remains untapped \citep[for non-VR approaches see][]{cherrueau2018enosstack,almaatouq2020empirica}.

\subsection{Data collection}

Most previous work in VR has focused on the \emph{data collection} and in particular the experimental content.
Second generation frameworks offered solutions for the technical implementation of the experiment by providing tasks, questionnaires, scores, and state machine (see Fig.~\ref{fig:six:pillars} Col. 3).
Frameworks automate many tedious and repetitive steps that have been plaguing VR studies by addressing the design-implementation-link problem.
Transforming an experiment protocol into a VR experiment is non-trivial.
Game engines a priori do not support experimental designs and therefore a mapping from a protocol to an implementation is required.
On the one hand, such a mapping can offer fine-grain control but requires more technical knowledge.
On the other hand, a coarse-grain control may not provide sufficient complex features to implement an experimental design.

Thus, there is a trade-off between a generalization of the experiment flow that allows to execute arbitrarily complexity \cite[e.g.][]{moulec2017agent, brookes2019studying} and simple plug-and-play approaches \cite[e.g.][]{grubel2016eve, vasser2017vrex, ugwitz2021toggle}.
Generalization usually provides researchers with a lot of control and at the same time the responsibility to write accurate experimental flows represented as finite-state-machines.
Plug-and-play approaches usually provide some prefabricated components that can be placed in the experiment and partially configured with configuration files. 
A middle ground can be found where experts can develop plug-and-play components based on a generalizable foundation that new users can easily use and with more experience start to adapt to their use case \cite[e.g.][]{grubel2016eve,brookes2019studying,starrett2020landmarks}.

Another important aspect is the Experiment Integration into background services.
Here it needs to be defined what data is to be collected, how the experiment can be managed while it is running, and how external content can be accessed.
Data collection requires a concept of the kind of data collected.
We can differentiate between virtual sensor data (collected in VR) and physical sensor data  \cite[collected by hardware, e.g., HMDs, physiological sensors;][]{grubel2016eve}.
Most frameworks focus on providing a data collection interface for virtual sensors in the virtual environment, disregarding signals from physical sensors, or only looking at virtualized information such as the position and orientation of HMDs and HIDs.

Especially in immersive VR, tracking the user action such as inputs can be difficult and requires special attention to ensure an accurate reconstruction \citep{howie2020virtual}.
Beyond measuring participants performance in VR, it is often useful to measure their attitudes with surveys \citep{grubel2016eve,tcha2016questionnaire}\citep{grubel2016eve} for subjective measures and if possible their physiological reaction \citep{weibel2018virtual} to underpin the mechanism investigated in a study beyond self-report and task completion.
Keeping only pre-analyzed results such as orientation error is argued to simplify research \citep{starrett2020landmarks}, but this does not account for calculation errors as occurred in some research projects requiring a redo.
Keeping all possible raw data is imperative to produce reproducible and auditable experiments.

Monitoring and managing experiment participants in real-time requires a definition and API of how the researcher can interact with the experiment as it happens.
Only two frameworks offer out of the box solutions for monitoring \citep{zhao2018networked,bebko2020bmltux}.
Especially, when studying crowd behavior, it may not be enough to conduct single-user experiments, but it becomes necessary to immerse multiple users at once requiring additional infrastructure and experiment integration \citep{moussaid2018virtual, zhao2018networked}.

\subsection{Data assembly}
The \emph{data assembly} is mostly secondary in nature in previous research.
Many frameworks opt for simple text files \cite[e.g.][]{zhao2018networked,starrett2020landmarks,watson2019use} to reduce the setup requirements for experiments.
Only a few frameworks opt for databases \cite[i.e.][]{grubel2016eve,tiwari2021define,howie2020virtual,wang2020vr,brookes2019studying} because they have more complex use cases in mind.
Large amounts of raw data from different sources make text files unwielding and require more effective data assembly.
Physical sensor data may be collected at a high frequency \citep{weibel2018virtual,watson2019use}, which would require special handling that is only integrated in data analysis or as a low frequency approximation during runtime to provide features such as scores.
In the best case, the complexity of data assembly can be hidden from the researchers' view \cite[e.g.][]{grubel2016eve} by only exposing the researcher to data collection and data analysis.

When running experiments online, with multiple participants or at multiple sites it becomes important to have remote data storage to effectively assemble data from multiple sources \cite[e.g.][]{zhao2018networked, brookes2019studying}.
Remote storage is orthogonal to questions of data representation in the assembly but requires infrastructure to transmit, store, and access data in the assembly later on.
A data serving API is required to organize remote storage and data representation.

Another aspect that has not been addressed by previous research is to transparently wrangle data.
Steps of anonymization, cleansing and reduction should be documented and reproducible.
In the context of data reduction it is possible to resolve, perform pre-analysis computations that others perform during data collection without maintaining raw data \citep{starrett2020landmarks}.

\subsection{Data analysis}
The \emph{data analysis} is usually not considered by second generation frameworks \citep[but see][]{grubel2016eve, schneider2018vreval,starrett2020landmarks,wang2020vr}.
The general move towards pre-registration makes integrating data analysis into frameworks a key requirement to enable researchers to actually go through with pre-registration without a massive overhead.
Some frameworks advocate to precompute task scores without keeping the raw data \citep{starrett2020landmarks} and present this as an advantage over post-processing by exposing the researcher to less tedious tasks.
However, in terms of auditability, having insight into the raw data is an advantage and if data analysis is integrated into the framework, then post-processing can also be largely automated and formally documented.

Effective data processing is crucial to scientific endeavours, but usually even basic checks are not reported.
Ensuring data quality is often not performed or only informally addressed by using robust statistical methods.
Data is only visualized based on summary metrics such as walked distance and time spent in spatial cognition experiments.
This contrasts with the richness of data that a VR experiment allows to be collected.
Providing visualisation tools, diagnostic tools for statistical properties, and other things would ensure that the data collected can be used for statistical analysis.

Obtaining insights from data is often performed independent of the experiment and the VR framework.
The advantage of keeping data insights within the framework is to tightly connect the analysis with the data collection mechanism and to greatly enhance the reproducibility and reusability.
It also allows simulations and machine learning analysis to rely on any collected data. 

\subsection{Management}
The \emph{Management} of an experiment is usually not codified in published research and research tools but implicitly performed \citep[but see][]{zhao2018networked,bebko2020bmltux}.
Management can include anything from participant acquisition, keeping an eye on the infrastructure status, to monitoring the overall experiment progress or using a live view of the participants' progress.
Some parts of the management of the experiments  may be reported, such as how participants where selected, but other duties are usually left out.
It is important to give special care to management as this neglected component of an experiment is often decisive for the quality of the outcome and the ability to reproduce the work at all.

\section{Frameworks for experimentation in VR through the lens of ExaC}
\label{app:frameworks:vr}

The frameworks summarized in Table \ref{tab:vr:frameworks:comparison} are discussed here in more detail to highlight each framework's contribution to the overall development of experiments in VR.
A focus on the six pillars of ExaC (see Fig.~\ref{fig:six:pillars} is maintained throughout the review to link the second generation frameworks potential for automation to aspects of third generation frameworks.

The Experiments in Virtual Environments (EVE) framework \citep{grubel2016eve} introduced a focus on automating data collection, assembly and analysis.
An experiment protocol is implemented using prefabricated (prefab) templates or creating new scenes with tasks and annotations on which data to store.
The design-implementation-link problem for novices is addressed with prefabs but experts are required to create new solutions.
At runtime, the framework manages the participant sessions and data assembly based on configuration files.
This framework also offers questionnaires and data analysis in an R package to bring together all data associated with an experiment in a single place and simplify difficult steps in data assembly and data analysis.
This setup allows for rapid prototyping and piloting including the analysis of the experiment, which is crucial for preregistration.
The framework is open-source and has been regularly extended with new templates based on studies conducted over the years \citep{weibel2018virtual, hackman2019neighborhood, fischer2020investigating, spielhofer2020physiological} and extended to run on all major operating systems.

The Virtual Reality Experiments (VREX)\citep{vasser2017vrex} was developed at the same time as EVE.
Plug-and-play design of experiments is fully achieved by providing a stand-alone software to edit the experiments based on Unity.
This limits the experiments to predefined tasks but in a fully configurable indoor environment.
Other interesting features are configurations for spatial audio and different locomotion systems. The data collection is a rudimentary data logging to files of fixed variables.
The authors also provide the actual unity project which would allow to create new tasks to expand the stand-alone version.

Expanding on the idea to automate difficult steps in experimental design in VR, the Automatic Generation of Experimental Protocol Run Time (AGENT) \citep{moulec2017agent} uses Domain-Specific Language (DSL) to facilitate the description of data collection. This approach is meant to reduce even further the workload on the researcher by generalizing typical experimental protocols. This approach works well in organizing the experimental protocol and generating the code to execute it.
However, it requires the researcher to link the generated code to objects in their scenes (design-implementation-link problem). Ultimately, the DSL cannot overcome the necessity to have in-depth technical knowledge of the game engine. AGENT also presumes the researcher to have a high-quality protocol in order to automatically generate the experiment code. As an external library, AGENT could be integrated with any game engine but only provides Unity API.

The framework VRate \citep{regal2018vrate} focuses on providing questionnaires in VR.
Questionnaires were implemented to work in immersive VR and are created from JavaScript Object Notations (JSON) and stored in comma separated (csv)files.
There is no support for the implemention of experiments beyond questionnaires.

Previous frameworks have focused on tailored environments to study general phenomena.
The VREVAL framework \citep{schneider2018vreval} focuses on automating experiment environments by integrating building information models (BIM).
Here, BIMs are combined with questionnaires, navigation tasks, and annotation tasks to conduct research that can be evaluated with external BIM software.
It is one of the few frameworks, to our knowledge, that uses the Unreal game engine.

Another interesting angle of research that is enabled by running VR studies is to understand how people react in crowds both under normal and emergency conditions. 
The Networked Virtual Reality for the Decision Science Laboratory (NVR-DeSciL) \citep{zhao2018networked} focuses on studying crowd dynamics in VR and is a networked solution to enable multiple participants in a single virtual environment at the same time.
An authoritative server architecture is used to provide the same experience to all participants.
Here, all user inputs are sent to the server and an update is produced for all participants at once.
The framework is tailored to the Decision Science Laboratory (DeSciL) at ETH Z\"urich but offers an interesting approach how to automate VR for multi-participant experiments.

Another framework that focuses on the execution of the experiment protocol is the Unity Experiment Framework (UXF) \citep{brookes2019studying}. Instead of using a DSL like AGENT, they programmatically separate the what from the how to address the design-implementation-link problem.
Here, the framework provides a generalization of the experiment flow in the form of sessions and trials and researchers must register with the sessions and trials what should happen in C\# code.
In a second step, the researcher must input how by writing C\# code snippets on how a trial within a session is run.
UXF provides many prebuild code snippets to design classic experiments.
While the setup does not quite follow  a classical ``\underline{\hspace{0.5cm}} as Code'' approach in the form of a configuration file, it can be considered one as the C\# code is essentially reading the configuration and executing it. 

The Unified Suite for Experiments (USE) \citep{watson2019use} focuses on neuroscience application by providing hard- and software to automate synchronizing data with high temporal precision in local labs.
The frame rate as well as transmission delays produce serious issues for brain-scanning where the exact stimulus must be determined.
USE introduces the SyncBox as a hardware setup that measures the current frame on the screen and aligns it ex-post with the high-frequency data from brain-scanners and other hardware.
The framework also provides a nested hierarchy and a user-defined experimental flows to create experiments tasks, providing yet another attempt how to solve the design-implementation-link problem. 
Additionally, the framework is able to automate interactions for artificial and non-human users.

Another attempt at the design-implementation-link problem has been taken by the framework Biomotion Lab Toolkit for Unity Experiments (bmlTUX) \citep{bebko2020bmltux}. It is a small framework that focuses on organizing participants into a factorial design with dependent and independent variables, providing a nested design for experiments but ultimately requires the coding of the trial. It compares to UXF or AGENT in providing a user interface for the factorial design instead of coding it in C\# or a DSL.

OpenMaze \citep{alsbury2020openmaze} focuses on a particular type of navigation experiments and tries to provide automations with maximal flexibility and low complexity.
Configuration files are used to define the relations between goals, landmarks, and enclosures. Here, the design-implementation-link problem is solved by restricting the input space and providing pre-configured implementations.

Data assembly is rarely the focus of a framework.
Virtual Observations \citep{howie2020virtual} focuses on the reconstruction of user actions.
In particular, immersive VR user input requires special care as handheld controllers spatially located in the virtual scene produce a large amount of data that critically defines the participants' action in the scene.
The framework advances replay and monitoring functionality but does not provide features for experimental design beyond recording the necessary data.
This framework is mostly orthogonal to other frameworks and could be used in parallel but without specific integration would duplicate data recording features. Nonetheless, the more fine-grained approach to user input in immersive VR is a valuable addition to automation practices.

The Landmarks framework \citep{starrett2020landmarks} focuses on data collection, assembly and analysis like EVE but sets different focuses. It also includes different immersions (HMD, Desktop, etc), the timeline of an experiment (a nested hierarchy) including tasks, the environment and the data collection with log-files. A unity package is provided easing the access to developing an experiment.

VR-Rides \citep{wang2020vr} is not a classical experiment framework but focuses on exercise games (exergames) where users physically move with either a pedaling device or a treadmill through a Google Street View (GSV). 
It provides automation for hardware and GSV.
Interestingly, the framework provides a user-study module that allows to assign participants to conditions, as well as a database allowing to implement experiments.

The Toggle Toolkit \citep{ugwitz2021toggle} provides an interesting approach to resolve the design-implementation-link problem.
It introduces \emph{trigger} as a design concept to transfer unity features to experimental design paradigms (autonomous trigger, key trigger, timer trigger, collider trigger, collider-key trigger, distance trigger).
The triggers can be chained producing something similar to a experimental flow but embedded in the scene.
While expressing complex task designs may be tedious, it can be performed without writing code providing an effective solution to the design-implementation-link problem.

The framework Delayed Feedback based Immersive Navigation Environment (DeFINE) for Studying Goal-Directed Human Navigation is an extension of UXF with a focus on the stimulus–response–feedback architecture for navigation experiments.
In particular, the feedback is an explicit part of the framework where other frameworks require the researcher to develop these features themselves.
The feedback is presented in the form of a leader board to encourage participants to perform better.
The framework also provides researchers with a set of locomotion methods including teleport, arm swing, head-bob and physical walking.
Lastly, it is also one of the few frameworks to provide immersed questionnaires to maintain immersion for participants.

\section{VR-Check for general VR experiment}
\label{app:VR:check}

The VR-Check \citep{krohn2020multidimensional} has been developed to underpin the development of VR software for medical use with the main purpose to retrain humans after severe physical or neuronal trauma.
While the original checklist is thus very domain specific, we generated a more general notion to help improve the development of an experimental protocol, experimental documentation and most of all, experimental design.
We rephrase the checklist as questions to make them more accessible to the researcher using them.
\begin{enumerate}
    \item \emph{Domain specificity}: Can the task in question produce a valid measurement for the research question?
    \item \emph{Ecological validity}: Does the measurement generalize to phenomena in the real world?
    \item \emph{Technical feasibility}: Can the task be measured with the given technical setup?
    \item \emph{User feasibility}: Can the users complete the task given the task  difficulty and the technical implementation (e.g. motion sickness)?
    \item \emph{User motivation}: Does the participant want to solve the task to the best of his ability?
    \item \emph{Task adaptability}: How fine-grained can the difficulty of the task be adjusted and can variations of the task be easily created?
    \item \emph{Performance quantification}: How suitable is the measurement to detect variation in task performance?
    \item \emph{Immersive capacities}: How does immersion (ability to perform the task based on hardware) compare to presence (the sense of being there influences task performance/user engagement) in the experimental design \citep[for a detailed discussion see][]{slater2009place,slater2016enhancing,slater2018immersion}?
    \item \emph{Training feasibility}: How often can the task be repeated to become solvable and how often has the task to be repeated to become trivial?
    \item \emph{Predictable pitfalls}: Where does the experimental design invite participants to behave other than intended?
\end{enumerate}


\bibliographystyle{spbasic}      
\bibliography{ref}   
